\documentclass{llncs}
\usepackage{makeidx}  
\usepackage{url}
\usepackage{graphicx}
\usepackage{multirow}
\usepackage{graphics}
\usepackage{epsfig}
\usepackage{psfrag}
\usepackage{multirow}
\usepackage{amsmath}
\usepackage{amsfonts}
\usepackage{mathtools}
\usepackage{caption}
\usepackage{subcaption}
\usepackage[breaklinks=true]{hyperref}
\usepackage{breakurl}

\makeatletter
\def\url@leostyle{%
 \@ifundefined{selectfont}{\def\UrlFont{\sf}}{\def\UrlFont{\small\ttfamily}}}
\makeatother
\DeclareMathOperator*{\argmin}{argmin}
\DeclareMathOperator*{\argmax}{argmax}

\graphicspath{{img/}{figs/}}

\begin{document}
\frontmatter          
\pagestyle{empty}  
\mainmatter              

\title {Learning a sparse database for patch-based medical image segmentation}
\titlerunning{Learning a sparse database for patch-based medical image segmentation}
 
\author{M. Freiman\inst{1} \and H. Nickisch\inst{2}  \and H. Schmitt\inst{2}  \and P. Maurovich-Horvat\inst{3} \and P. Donnelly\inst{4} \and M. Vembar\inst{5}  \and L. Goshen\inst{1}  }
 \authorrunning{M. Freiman et al.}   
\tocauthor{M. Freiman (GAT, CT, Philips Healthcare, Haifa, Israel.),
H. Nickisch (Philips Research, Hamburg, Germany.),
H. Schmitt (Philips Research, Hamburg, Germany.),
P. Maurovich-Horvat (Heart and Vascular Center,Semmelweis University, Budapest, Hungary)
P. Donnelly (South Eastern Health and Social Care Trust, Queen's University, Belfast, Ireland)
M. Vembar (Clinical Science, CT, Philips Healthcare, Cleveland, Ohio, USA),
L. Goshen (GAT, CT, Philips Healthcare, Haifa, Israel.)
 }
\institute{GAT, CT, Philips Healthcare, Haifa, Israel. \and Philips Research, Hamburg, Germany \and Heart and Vascular Center,
Semmelweis University, Budapest, Hungary \and South Eastern Health and Social Care Trust, Queen's University, Belfast, Ireland \and Clinical Science, CT, Philips Healthcare, Cleveland, Ohio, USA
 %
 \newline
 \email{moti.freiman@philips.com}
 }
%


\maketitle              

\begin{abstract}
We introduce a functional for the learning of an optimal database for patch-based image segmentation with application to coronary lumen segmentation from coronary computed tomography angiography (CCTA) data. The proposed functional consists of fidelity, sparseness and robustness to small-variations terms and their associated weights. Existing work address database optimization by prototype selection aiming to optimize the database by either adding or removing prototypes according to a set of predefined rules. In contrast, we formulate the database optimization task as an energy minimization problem that can be solved using standard numerical tools.
We apply the proposed database optimization functional to the task of optimizing a database for patch-base coronary lumen segmentation. Our experiments using the publicly available MICCAI 2012 coronary lumen segmentation challenge data show that optimizing the database using the proposed approach reduced database size by 96\% while maintaining the same level of lumen segmentation accuracy. Moreover, we show that the optimized database yields an improved specificity of CCTA based fractional flow reserve (0.73 vs 0.7 for all lesions and 0.68 vs 0.65 for obstructive lesions) using a training set of 132 (76 obstructive) coronary lesions with invasively measured FFR as the reference.

\end{abstract}

\section{Introduction}
\label{sec:intro}
Segmentation of anatomical structures from medical images plays an important role in many clinical applications. Automatic segmentation can be challenging due to the large variability in anatomical structures shape and appearance. Patch-based, non-parametric segmentation algorithms such as the K-nearest neighbor (KNN) algorithm \cite{Cover1967} have demonstrated their potential in automatic segmentation of challenging anatomical structures. For example, Mechrez et al. \cite{Mechrez2016} use the KNN algorithm followed by a spatial consistency refinement step to segment Multiple-Sclerosis lesions from MRI data, and Wang et al. \cite{Wang2014} demonstrate the potential of KNN algorithm in defining a search-space for improved patch-based segmentation of cardiac MR data and abdominal CT data. Specifically in the cardiovascular domain, Olabarriaga et al \cite{Olabarriaga2005} used the KNN algorithm to steer a model-based segmentation of abdominal aortic aneurysms and more recently, Freiman et al. \cite{Freiman2017} used the KNN algorithm to estimate the likelihood component within a graph min-cut framework for coronary artery lumen segmentation. 

While the KNN algorithm has several theoretical and practical advantages \cite{Cover1967}, two main limitations of this algorithm are: 1) large storage to retain the set of examples which defines the training set, and 2) low efficiency due to the re-calculation of the similarity between the test and training samples at each evaluation \cite{Garcia2012}.

Among the approaches previously proposed to address these issues, database optimization by prototype selection is an attractive approach as it maintains originally an-notated data rather than generating artificial data. The optimal prototype selection is an NP-hard problem which can be mapped onto a set-cover problem and solved using an approximation algorithm \cite{Bien2011}. Alternatively, the prototype selection problem can be relaxed by introducing some order on the prototypes.  Then, the prototype selection approaches can be divided into three categories: 1) incremental search, in which the algorithm adds prototypes to the reduced data based on some rule, 2) decremental search in which the algorithm aims to remove prototypes from the database according to some rule, and; 3) hybrid approach which combines both incremental and decremental steps. For a comprehensive review of methods aimed to reduce KNN algorithm storage and computational demand we refer the reader to Garc\'{i}a et al \cite{Garcia2012}. 

In this work we formulate the database optimization problem as an energy minimization which enables the optimization using common numerical approaches. Our functional consists of fidelity, sparsity and robustness to small variations terms along with their associated weights.  We applied the proposed functional to optimize a database used in Freiman's et al patch-based coronary artery lumen segmentation algorithm \cite{Freiman2017}. We evaluated the influence of database optimization on the segmentation performance by means of segmentation accuracy using the publicly available MICCAI 2012 Coronary Lumen Segmentation Challenge Database \cite{Kirisli2013}. We also evaluated the impact of the database optimization on CCTA based fractional flow reserve (CT-FFR) estimates using a database of 132 coronary lesions with invasively measured FFR as the reference.

Our results show that the database size can be reduced by 96\% while maintaining the same level of coronary lumen segmentation accuracy on the MICCAI 2012 Coronary lumen segmentation challenge database \cite{Kirisli2013} and even improving the performance of CT-FFR estimates obtained with 3D models generated from the automatic segmentation results.

\section{Method}
\label{sec:method}
Our goal is to select a subset of prototypes from a given database so that the classification performance for any new sample will be as accurate as possible. The sub-sampled database should represent the full structure of the population with as few prototypes as possible. 
First, we define a property describing the distribution of the prototypes in the original database. Next, we define a set of parameters used to generate a sub-sampled database, and finally, we formulate the optimal parameter finding as an energy minimization problem.

\subsection{Prototype ranking}

Inspired by the work of Bein and Tibshirani \cite{Bien2011}, we rank prototypes in the original database according to their location on the manifold. Specifically, we will consider a prototype as located in the center of its class when its neighboring most similar prototypes according to some pre-defined metric are from of the same class and as located on the boundary between classes if it has many neighbors similar samples which are belonging to other  from different classes. Formally,
for a  prototype feature-vector $\vec{x}$, we define the sample ranking score $R(\vec{x})$ as follows: 
\begin{equation}
\label{eq:rank}
    R(\vec{x}) = \frac{\text{\#NN with other class}}{\text{\#NN with same class}} 
               = \frac{          \sum_{k=1}^K 1-\delta(C[\vec{x}],C[\vec{x_k}])} {\text{max}\left(\sum_{k=1}^K  \delta(C[\vec{x}],C[\vec{x_k}]),1\right)} 
\end{equation}
where $K$ is the number of the nearest prototype neighbors $\vec{x_i}$ that are similar according to the chosen distance metric, $C[\vec{x_i}]$ is the class of $\vec{x_i}$, $\vec{x}$ is a prototype similar to $\vec{x_i}$, and: 
\begin{equation}
\label{eq:C}
\delta(C[\vec{x_i}],C[\vec{x_k}]) = \begin{cases}
    1,& C[\vec{x_i}]=    C[\vec{x_k}] \\
    0,& C[\vec{x_i}]\neq C[\vec{x_k}]
\end{cases}
\end{equation}
According to this definition $R(\vec{x})$ gets a
high value when the prototype $\vec{x}$ has many similar prototypes from other classes, and a low value when the prototype $\vec{x}$ has many similar prototypes from its own class.

\subsection{Database sparsification}

We describe the distribution of the classification sample ranking scores of the prototypes in the original database for each class using a histogram with $N$ bins. The bins boundaries are set (and kept fixed) to be percentiles of the overall samples per class to normalize against various sample ranks distributions. We define a sparsified database as a function of the percentile of prototypes to be selected from each bin of the histogram $DB(\vec{N})$, where $\vec{N}\in\mathbb{N}^N$ is the vector containing the number of prototypes to be put in each of the $N$ bins. The subsampling is done deterministically by selecting the $N_i$ prototypes with highest ranking score for bin $i$.

\subsection{Database optimization}

Given the function $DB(\vec{N})$ to sample the original database, we define a functional to estimate the sampling parameters $\vec{N}$ i.e. the number of prototypes per bin.  Our goal is to find $\vec{N}$ that maximizes the capability of the sampled database to correctly classify each sample while minimizing the overall number of samples. We also would like the classification to be robust to small variations in the samples. Formally, our functional is defined as:    

\begin{equation}
\label{eq:energy_func}
E_{\alpha,\beta}(\vec{N}) = \sum_{i=1}^{M} \overbrace{ \rho(\vec{N},\vec{x_i})}^{\text{robustness}}
  + \alpha \sum_{i=1}^{M} \overbrace{ \left(C(\vec{x_i})-f\left(DB(\vec{N}),\vec{x_i}\right)\right)^2 }^{\text{fidelity}}
  + \beta \overbrace{\|DB(\vec{N})\|}^{\text{sparsity}}
\end{equation}
where: $DB(\vec{N})$ is the sampled database constructed from the original database by sampling the different bins according to the percentiles specified by $\vec{N}$, $f\left(DB(\vec{N}),\vec{x_i}\right)$ is the classification of the prototype $\vec{x_i}$ using the sampled database $DB(\vec{N})$, $\|DB(\vec{N})\|$ is the number of the prototype in the sampled database $DB(\vec{N})$, $M$ is the number of prototypes in the original database, $\alpha$,$\beta$  are weighting meta-parameters controlling the contribution of  each term, and $TV(\vec{N})$ measures the robustness of the the classification of each prototype using $DB(\vec{N})$ to a small variation in its appearance as follows:
\begin{equation}
\label{eq:TV}
\rho(\vec{N},\vec{x})  = \sum_{j=1}^{J} \|f\left(DB(\vec{N}),\vec{x}\right)-f\left(DB(\vec{N}),\vec{x}+\vec{e_j}\right)\|_1
\end{equation}
where $J$ is the dimension of the prototype $\vec{x}$, the unit vector $\vec{e_j}$ is of the same dimension as with zeros at all entries except at entry $j$, and $\|f\left(DB(\vec{N}),\vec{x}\right)-f\left(DB(\vec{N}),\vec{x}+\vec{e_j}\right)\|_1$ is the absolute difference between the classification of $\vec{x}$ and $\vec{x}+\vec{e_j}$ given the database $DB(\vec{N})$.

We find the optimal database sampling parameters by minimizing the energy functional:
\begin{equation}
\label{eq:energy_min}
\hat{\vec{N}} = \argmin_{\vec{N}} E_{\alpha,\beta}(\vec{N})
\end{equation}

\subsection{Coronary Lumen Segmentation}
\label{subsec:cls}

We apply the proposed approach to reduce the database size required for the coronary lumen segmentation algorithm proposed by Freiman et al. \cite{Freiman2017}. For the sake of completeness, we briefly describe the relevant parts of the algorithm here. We refer the interested reader to a detailed and complete description provided in \cite{Freiman2017}. 
The algorithm formulates the segmentation task as an energy minimization problem over a cylindrical coordinate system \cite{Lugauer2014} where the warped volume along the coronary artery centerline is expressed with the coordinate $i$ representing the index of the cross-sectional plane, and $\theta$, $r$ represent the angle and the radial distance determining a point in the cross-sectional plane
\begin{equation}
	E(X)=\sum_{p \in P} \psi_p(x_p) +\lambda \sum_{p,q \in E} \varphi_{p,q}(x_p,x_q)
\end{equation} 
where $P$ is the set of sampled points, $x_p$ is a vertex in the graph representing the point $(i^{x_p},\theta^{x_p},r^{x_p})$ sampled from the original CCTA volume, $\psi_p(x_p)$ represents the likelihood of the vertex to belong to the lumen or the background class, $p$, $q$ are neighboring vertices according to the employed neighboring system $E$, and $\varphi_{p,q}(x_p,x_q)$ is a penalty for neighboring vertices belonging to different classes ensure the smoothness of the resulted surface.
The algorithm uses the KNN algorithm \cite{Cover1967} to calculate the likelihood of each vertex $x_p$ belonging to the coronary lumen from a large training database with rays sampled from cardiac CTA data along with manually edited lumen boundary location  represented as a  binary rays serving as the database prototypes.
The likelihood term is additionally adjusted to account for partial volume effects and a $L2$ norm used as the regularization term as described in \cite{Freiman2017}.

\subsection{Application of the database optimization}
The original database used in the coronary lumen segmentation algorithm consists of $\sim$2,130,000 prototypes obtained from 97 CCTA datasets segmented manually by a cardiac CT expert. We consider the lumen radii as the different classes of the rays. We use the L2 norm as the distance metric to define the sample rank (Eq.~\ref{eq:rank}), and experimentally set the number of bins in the histogram N (Eq.~\ref{eq:energy_func}) to 5.  We optimize the functional hyperparameters to achieve the maximal area under the curve (AUC) for CT-FFR estimates with the segmentations obtained using the optimized database with invasive FFR measurements as the reference. Formally, we define a two-phase optimization task, where the outer loop optimized the model hyperparameters $\alpha$ and $\beta$:
\begin{equation}
\label{eq:hyper_min}
	\widehat{\alpha,\beta} = \argmax_{\alpha,\beta} AUC \left(FFR_{CT} \left(DB\left(\vec{N}\right)\right) - FFR_{GT}\right)  
\end{equation}
and the inner loop find the optimal model parameters for given $\alpha$,
$\beta$ using Eq.~\ref{eq:energy_min}.
We carried out the optimization using the derivative-free BOBYQA algorithm by Powel et al. \cite{Powell2009}.

\section{Experimental results}
\label{sec:exp_results}
We use two data sets as follows. The first dataset consists of CCTA data of 132 coronary lesions that were retrospectively collected from the medical records of 97 subjects who underwent a CCTA and invasive coronary angiography with invasive FFR measurements due to suspected CAD. CCTA data was acquired using either a Philips Brilliance iCT (gantry rotation time of 0.27 sec.) or Philips Brilliance 64 (gantry rotation time of 0.42 sec.). Acquisition mode was either helical retrospective ECG gating (N=54) or prospectively ECG triggered axial scan (N=43). The kVp range was 80 -- 140 kVp and the tube output range was 600 -- 1000 mAs for the helical retrospective scans and 200 -- 300 mAs for the prospectively ECG triggered scans. 

Cross-sectional area (CSA) based stenosis quantification was performed by an expert reader on 132 lesions, out of which 56 were diagnosed as non-obstructive lesions (CSA stenosis less than 50\%) and 76 diagnosed as obstructive lesions (CSA stenosis 50\%-90\%).  According to the invasive FFR measurements, 48 lesions were hemodynamically significant (FFR$\leq$0.8) and 84 lesions were non-significant (FFR$>$0.8).  
The coronary artery centerlines and the aorta segmentation were computed automatically and adjusted manually by a cardiac CT expert (--) to account for algorithm inaccuracies using a commercially available software dedicated to cardiac image analysis (Comprehensive Cardiac Analysis, IntelliSpace Portal 6.0, Philips Healthcare).

We tuned the model hyper-parameters and optimized the database using Eq.~\ref{eq:energy_min} and ~\ref{eq:hyper_min}. Next, we segmented the coronary tree with Freiman's et al algorithm \cite{Freiman2017} with the full and the optimized databases. Finally, we performed the flow simulations using the lumped parameter model (LM) proposed by Nickisch et al. \cite{Nickisch2015}.

\begin{figure}[t!]
\centering
		\psfrag{Energy minimization based database dilution}[cc][][2.5][0]{Energy minimization based database dilution}
\psfrag{percent}[cc][][2.25][0]{$\%$}
\psfrag{Bin idx}[cc][][2.25][0]{Bin idx}

\psfrag{0.005}[cc][][1.2][0]{0.005}
\psfrag{0.195}[cc][][1.2][0]{0.195}
\psfrag{0.00687}[cc][][1.2][0]{0.00687}
\psfrag{0}[cc][][1.5][0]{0}
\psfrag{0.1}[cc][][1.5][0]{0.1}
\psfrag{0.2}[cc][][1.5][0]{0.2}
\psfrag{0.3}[cc][][1.5][0]{0.3}
\psfrag{0.4}[cc][][1.5][0]{0.4}
\psfrag{0.5}[cc][][1.5][0]{0.5}
\psfrag{0.6}[cc][][1.5][0]{0.6}
\psfrag{0.7}[cc][][1.5][0]{0.7}
\psfrag{0.8}[cc][][1.5][0]{0.8}
\psfrag{0.9}[cc][][1.5][0]{0.9}
\psfrag{1}[cc][][1.5][0]{1}
\psfrag{2}[cc][][1.5][0]{2}
\psfrag{3}[cc][][1.5][0]{3}
\psfrag{4}[cc][][1.5][0]{4}
\psfrag{5}[cc][][1.5][0]{5}
\resizebox{\columnwidth}{!}{\includegraphics{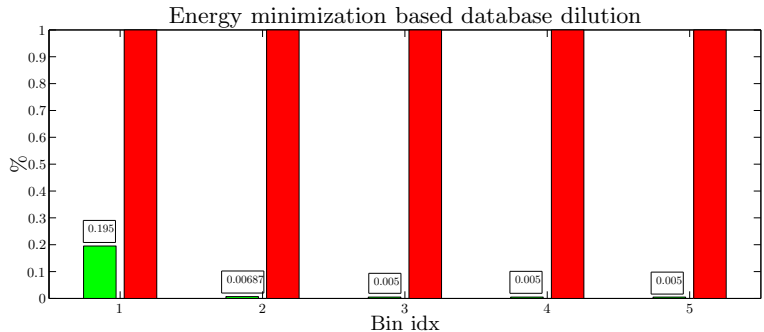}}
		
\caption{The database reduction for each bin of the histogram of prototypes. The red bars indicate 100\% of the original database and the green bars indicates the percentage of the remaining prototypes after the optimization. Optimized database sampling parameter for each bin is listed above the optimized histogram.}
\label{fig:sampling}
\end{figure}

We compared the flow simulation results from the 3D coronary tree models generated using the coronary lumen segmentation algorithm described in section~\ref{subsec:cls}, with the full and optimized training databases. 
The optimization process reduced the database size by 96\% from $\sim$2,130,000 prototypes to $\sim$84,000 prototypes. Fig.~\ref{fig:sampling} illustrates the reduction in the number of prototypes in the database at the different bins of the prototypes histogram. The Functional-based optimization prefers to keep prototypes with low sample ratio (i.e closer to the center of mass of each class) in the optimized database. Table~\ref{table:ffr_results} summarizes the performance metrics for assessing the hemodynamic significance of coronary lesions with automatic segmentation using the entire and optimized database for entire set of coronary lesions and specifically for obstructive lesions (Cross Sectional Area (CSA) stenosis$\geq$50\%). The flow simulation results are slightly better using the optimized database compared to the results of the full database, although the optimized database has much less prototypes compared to the full database.

\begin{table}[t!]
\centering
\caption{Summary statistics of hemodynamic significance assessment of coronary lesions by means of CT-FRR based on automatic segmentation with the entire database and with the optimized database. Results presented for all (N=132) lesions and specifically for obstructive (Obst.) lesions (N=76) separately.}
\label{table:ffr_results}
\vspace{1mm}
\begin{tabular}{|l|c|c|c|c|c|c|c|c|}
\hline
\multirow{2}{*}{} & \multicolumn{2}{c|}{\textbf{Sensitivity}} & \multicolumn{2}{c|}{\textbf{Specificity}} & \multicolumn{2}{c|}{\textbf{Accuracy}} & \multicolumn{2}{c|}{\textbf{AUC}} \\ \cline{2-9} 
 & \multicolumn{1}{c|}{All} & \multicolumn{1}{c|}{Obst.} & \multicolumn{1}{c|}{All} & \multicolumn{1}{c|}{Obst.} & \multicolumn{1}{c|}{All} & \multicolumn{1}{c|}{Obst.} & \multicolumn{1}{c|}{All} & \multicolumn{1}{c|}{Obst.} \\ \hline
\textbf{\begin{tabular}[c]{@{}l@{}}Optimized\\ database\end{tabular}} & 0.85 & 0.86 & 0.73 & 0.68 & 0.77 & 0.76 & 0.84 & 0.83 \\ \hline
\textbf{\begin{tabular}[c]{@{}l@{}}Full\\ database\end{tabular}} & 0.85 & 0.86 & 0.70 & 0.65 & 0.76 & 0.75 & 0.84 & 0.82 \\ \hline
\end{tabular}
\end{table}

\begin{figure}[t!]
\centering
\resizebox{\columnwidth}{!}{\includegraphics{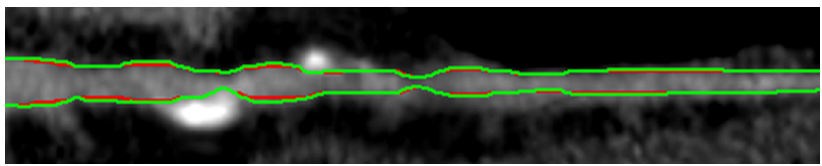}}
\caption{Representative example of straight multi-planar reconstructed images of coronary artery segmentation results with the optimized (green) and full database.}
\label{fig:vessel}
\end{figure}

Fig.~\ref{fig:vessel} depicts representative examples of straight multi-planar reconstructed images of coronary artery segmentation results with the optimized (green) and the full (red) database. Table~\ref{table:miccai_res} presents the segmentation accuracy results of our algorithm with the original and optimized database using the MICCAI 2012 challenge framework training data \cite{Kirisli2013}. We refer the reader to the challenge website \cite{Kirisli2013} for further comparison with the rest of the methods and with the observer performance.

\begin{table}[t!]
\centering
\caption{Summary statistics of coronary lumen segmentation accuracy using the MICCAI 2012 challenge evaluation framework \cite{Kirisli2013} for the training datasets (18 cases, 78 coronary segments). Results presented for healthy and diseased segments separately and in the relevant metric units.}
\label{table:miccai_res}
\vspace{1mm}
\begin{tabular}{|l|c|c|c|c|c|c|}
\hline
\multirow{2}{*}{} & \multicolumn{2}{c|}{\textbf{Dice (\%)}} & \multicolumn{2}{c|}{\textbf{MSD (mm)}} & \multicolumn{2}{c|}{\textbf{MAX SD (mm)}} \\ \cline{2-7} 
 & Healthy & Disease & Healthy & Disease & Healthy & Disease \\ \hline
\textbf{\begin{tabular}[c]{@{}l@{}}Optimized\\ database\end{tabular}} & 0.69 & 0.74 & 0.49 & 0.27 & 1.69 & 1.24 \\ \hline
\textbf{\begin{tabular}[c]{@{}l@{}}Full\\ database\end{tabular}} & 0.69 & 0.74 & 0.49 & 0.27 & 1.69 & 1.22 \\ \hline
\end{tabular}
\end{table}

\section{Conclusion}
\label{sec:conclusion}
We presented an energy functional for optimizing the training database in patch-based medical image segmentation algorithms. We define a 'sample rank' order on the training database prototypes and formulate the prototype sampling as an energy minimization task with hyper-parameters that can be adjusted to the specific task. 
We demonstrated the application of this approach to reducing database size and improving the performance of coronary lumen segmentation algorithm from CCTA data. Our experiments show that the optimized database can maintain overall segmentation results with added incremental improvements of CT-FFR estimates based on the 3D models generated from the segmentation results while substantially reducing the memory demand of the algorithm. 


\bibliographystyle{splncs03}
\bibliography{MICCAI2017-PATCHMI}

\end{document}